\documentclass[11pt,a4paper]{article}

\usepackage{epsfig}
\usepackage{latexsym}
\usepackage{graphicx}

\newcommand{\bi}{\begin{itemize}}
\newcommand{\ei}{\end{itemize}}
\newcommand{\be}{\begin{equation}}
\newcommand{\ee}{\end{equation}}

\newcommand{\bea}{\begin{eqnarray}}
\newcommand{\eea}{\end{eqnarray}}
\newcommand{\beastar}{\begin{eqnarray*}}
\newcommand{\eeastar}{\end{eqnarray*}}

\newcommand{\eq}[1]{~(\ref{#1})}

\newcommand{\CC}{{\cal C}}

\newcommand{\s}{\sigma}

\begin{document}

\title{Intermittency of glassy relaxation and the emergence
of a non-equilibrium spontaneous measure in the aging regime}

\author{A. Crisanti$^{\dag}$ and F Ritort$^{\ddag,\#}$\\
$\dag$ Dipartimento di Fisica, Universit\`a di Roma ``La
  Sapienza''\\  INFM Sezione di Roma I and SMC,
  P.le Aldo Moro 2, 00185 Roma, Italy\\
$\ddag$  Department of Physics, University of California, Berkeley CA 94720,
USA\\
$\#$ Department of Physics, Faculty of Physics,
University of Barcelona\\ Diagonal 647, 08028 Barcelona, Spain\\
%2) Department of Physics, University of California\\ Berkeley CA 94720,
%USA\\
{\tt Corresponding author:ritort@ffn.ub.es}}

\maketitle

\abstract{\bf We consider heat exchange processes between
non-equilibrium aging systems (in their activated regime) and the
thermal bath in contact. We discuss a scenario where two different
heat exchange processes concur in the overall heat dissipation: a
stimulated fast process determined by the temperature of the bath and
a spontaneous intermittent process determined by the fact that the
system has been prepared in a non-equilibrium state.  The latter is
described by a probability distribution function (PDF) that has an
exponential tail of width given by a parameter $\lambda$, and
satisfies a fluctuation theorem (FT) governed by that parameter.  The
value of $\lambda$ is proportional to the so-called effective
temperature, thereby providing a practical way to experimentally
measure it by analyzing the PDF of intermittent events.}

After decades of research many aspects of the glass state still defy
our comprehension. Mode-coupling theory (MCT) provides a consistent
framework~\cite{GotSjo92} to describe relaxational processes
experimentally observed in a given temperature range. However, the
structural arrest predicted by the ideal version of MCT is not
observed, thereby activated processes have been advocated to drive the
relaxation toward the equilibrium state. A physical description of
these activated processes remains obscure as no clear experimental
identification has been established. It has been long
suspected~\cite{AdaGib65} that activated processes in glasses are not
driven by free-energy differences between nucleating phases
(e.g. liquid versus solid) but rather by an entropic mechanism
involving decay among a large number of physically indistinguishable
coexisting phases.  This coexistence is a rather new concept in
statistical physics, however it is one of the central elements in
spin-glass theory~\cite{MezParVir87}, known to provide a good
qualitative description of many aspects experimentally observed in
disordered magnets and structural glasses.  From the experimental
point of view the identification of such activated processes appears
challenging as no specific physical property can be targeted. The
presently accepted view is that activated processes are experimentally
accessible through direct observation of cooperative motion of small
(nanometer sized) regions of particles, also called cooperatively
rearranging regions (CRRs). Indirect evidences of activation events
have been already reported in confocal scanning
microscopy~\cite{KegBla00,WeeCroLevSchWei00} and light
scattering~\cite{BisTraRomCip03} (using time resolved correlation
analysis) of colloidal systems.  In addition, dielectric noise
sensitive measurements in glass formers have detected intermittent
voltage signals with long tails in the corresponding probability
distribution functions (PDFs)~\cite{VidIsr98,Ciliberto03}.  The
purpose of this paper is to show that activated processes in
structural glasses are related to intermittency effects theoretically
described in terms of a spontaneous heat exchange process that
generates a non-equilibrium measure and an associated fluctuation
theorem (FT). Intermittency measurements provide a way to identify
these spontaneous processes as well as quantify FDT violations.

Let us consider a glass former quenched from a high temperature (where
the system is initially equilibrated) to a lower value $T< T_{\rm
MCT}$ ($T_{\rm MCT}$ being the MCT temperature) where slow relaxation
and activated dynamics sets in. During the aging process heat $Q$ is
constantly exchanged between the system and the bath.  According to
our view, this exchange occurs in two different ways that we will
refer as stimulated and spontaneous. The stimulated way corresponds to
a continuous back and forth heat exchange between the sample and the
bath, the net average heat exchange being zero on timescales much
smaller than the age of the system $t_w$. Were one to measure the
statistical distribution $P^{\rm st}(Q)$ of the heat released by this
process from system to bath along $N_t$ time intervals of regular
duration $\Delta t$, a Gaussian distribution centered around zero
would be observed. We call this process stimulated because it is
strictly dependent on the presence of the thermal bath meaning that
the variance of $P^{\rm st}(Q)$ is dependent on $T$ but independent of
$t_w$. The spontaneous way of relaxation is different as it occurs in
a much longer timescale compared to the stimulated way.  It occurs by
an intermittent heat release from the system to the surroundings
(and from there to the thermal bath), yet the net average of heat
supplied by this mechanism is not zero.  For were one to measure the
statistical distribution $P^{\rm sp}(Q)$ of the intermittent heat
released from system to the bath by this process an exponential tail,
characteristic of first order Markov processes, would be observed.
This release of heat we call spontaneous because it is determined by
the fact that the system has been prepared in a non-equilibrium state
rather than by the value of the temperature $T$ of the
bath. Therefore, in contrast to the previous case, the width of the
exponential tail would be $T$ independent but $t_w$ dependent and
gradually change as the final equilibrium state is approached. 
During an aging process both heat exchange mechanisms appear
intertwined in a complex pattern, yet it would be possible to discern
them by measuring the distribution $P(Q)$ of heat exchanged and
disentangling the Gaussian and exponential components. Unfortunately,
the measurement of heat transfer between the system and the bath
appears very difficult so one has to look for indirect ways of
detecting the existence of these two processes. We aim to  
validate the previous scenario by doing numerical simulations on
a given class of models of glass forming liquids and propose
some predictions to be challenged in experiments.

To discern the stimulated and the spontaneous components from the
global statistical distribution of transferred heat $P_{t_w}(Q)$, the
following protocol has been implemented (other choices lead to similar
results, yet the the present one has revealed the most efficient). Let
us consider systems with stochastic dynamics. Initially the system is
equilibrated at a temperature $T_i\gg T_{\rm MCT}$. Then the
temperature is instantaneously changed to $T<T_{\rm MCT}$ (this
defines the initial time $t=0$) and configurations evolved. In
addition to the runtime configuration it is convenient to keep track
of the ``valley'' to which a given configuration belongs. A
constructive approach to identify valleys has been proposed by
Stillinger and Weber who have considered a topographic view of the
potential energy landscape where valleys are called inherent
structures (IS)~\cite{StiWeb82}.  The system is aged for a time $t_w$
and the corresponding IS recorded.  Dynamics continues until a new IS
is found, thereby defining the {\em first jump}.  The exchanged heat
$Q$ corresponding to that jump is recorded.  For stochastic systems no
work is exerted upon the system so $Q=\Delta E=E(\CC')-E(\CC)$ where
$\CC$ is the runtime configuration at $t_w$ and $\CC'$ the runtime
configuration just after the first jump occurs. If $Q>0$ heat is
transferred from the bath toward the system and vice versa. The
quenching experiment is repeated many times, each quench a value of
$Q$ is obtained, and the probability distribution function (PDF)
$P_{t_w}(Q)$ measured. Qualitative identical results are obtained if
the runtime configuration is used to control when the first jump
occurs. The main advantage of keeping track of the IS is that it
provides a useful way to filter out collective spin (or particles)
rearrangements.  Related procedures have been used to analyze trapping
time distributions~\cite{BucHeu00,DenReiBou03,DalSib03}. Here we will
concentrate our attention in the exchange heat distribution, trying
to establish a link between the intermittency observed and the
existence of a FT describing the spontaneous process.

We have considered the random-orthogonal
spin-glass model (ROM)~\cite{MarParRit94} which deserves interest as
it is a good microscopic realization of the random-energy model, a
phenomenological model of inherent simplicity commonly used in the
study of disordered systems~\cite{MezParVir87}. Three different
reasons motivate our choice: 1) The ROM is of the mean-field type (as
interactions are long ranged) and shows dynamical properties in
agreement with the predictions of the ideal version of MCT. It has
been shown~\cite{MarParRit94} to have a MCT transition temperature
$T_{\rm MCT}$ below which the relaxation time diverges exponentially
fast with $N$; 2) If $T<T_{\rm MCT}$ and the number of spins $N$ is
not too large then activated processes are observable for long enough
times~\cite{AroBovGay02}. Moreover, activation barriers can be tuned
by changing the size of the system~\cite{CriRit00}; 3) Computation of
$P_{t_w}(Q)$ is numerically affordable.

The ROM is defined in terms of a set of $N$ spin variables that can
take two values $\s_i=\pm 1$, each configuration corresponding to a
set of spin values, $\CC=\lbrace\s_i;1\le i\le N \rbrace$. Spins
interact via random exchange couplings $J_{ij}$ leading to an energy
function with strong disorder-induced frustration and therefore to a
complex pattern of local minima. The energy $E$ of a configuration is
given by $E=-2\sum_{i<j}J_{ij}\s_i\s_j$ where the $J_{ij}$ are
Gaussian distributed variables (zero mean and variance $1/N$) with
correlations $\sum_{i=1}^NJ_{ij}J_{ik}=\delta_{jk}$. A Monte Carlo
dynamics is implemented where spins are randomly selected and updated
$\s_i\to -\s_i$ depending on the energy change $\Delta E$ according to
the Metropolis algorithm.  For sake of clarity, all along the paper,
we will present results for $N=64$ where the statistics collected is
much better. Nevertheless, as a check of the correctness of our
results, other sizes have been investigated $N=32,48,300$ finding
identical results in all cases (the $N$ dependence of the values of
the trapping times ensure that we are indeed observing activated
processes). The ROM has a MCT transition temperature $T_{\rm
MCT}\simeq 0.53$. Three quenching temperatures have been investigated
$T=0.3,0.2,0.1$ and around $10^4$ different quenches have been
collected for each experiment.

\begin{figure}
\begin{center}
\epsfig{file=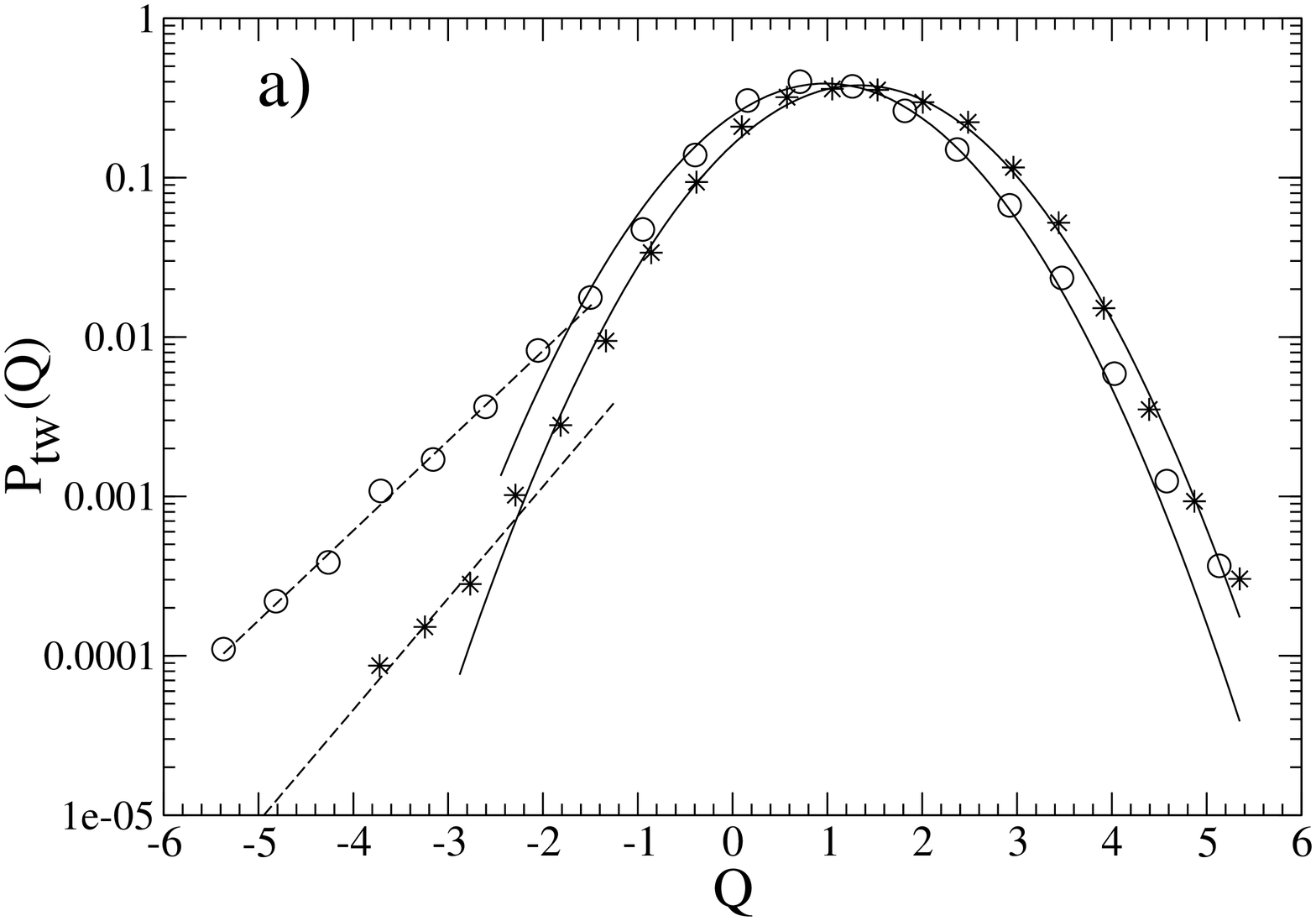,angle=-90,width=10cm}
\epsfig{file=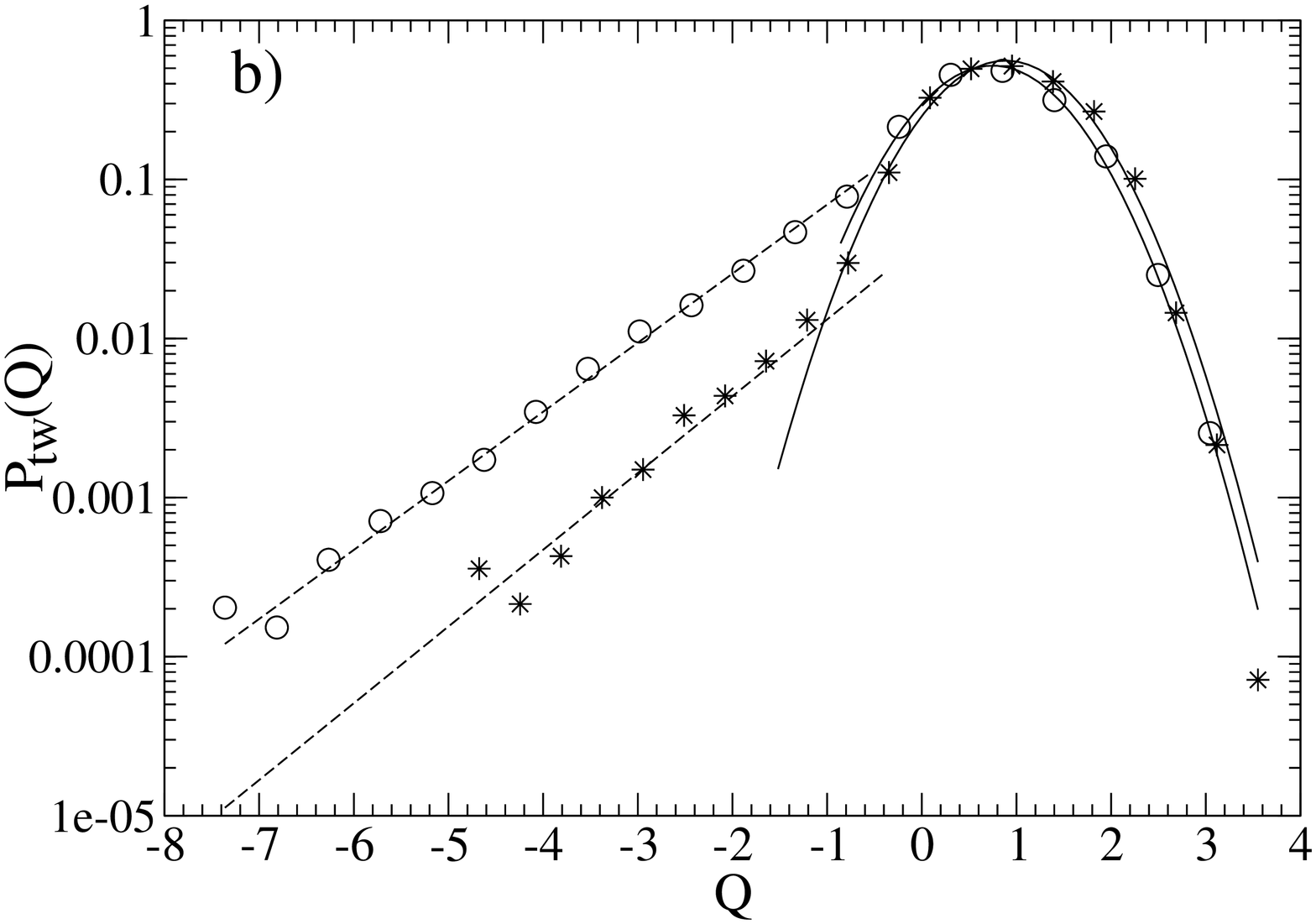,angle=-90,width=9cm}\epsfig{file=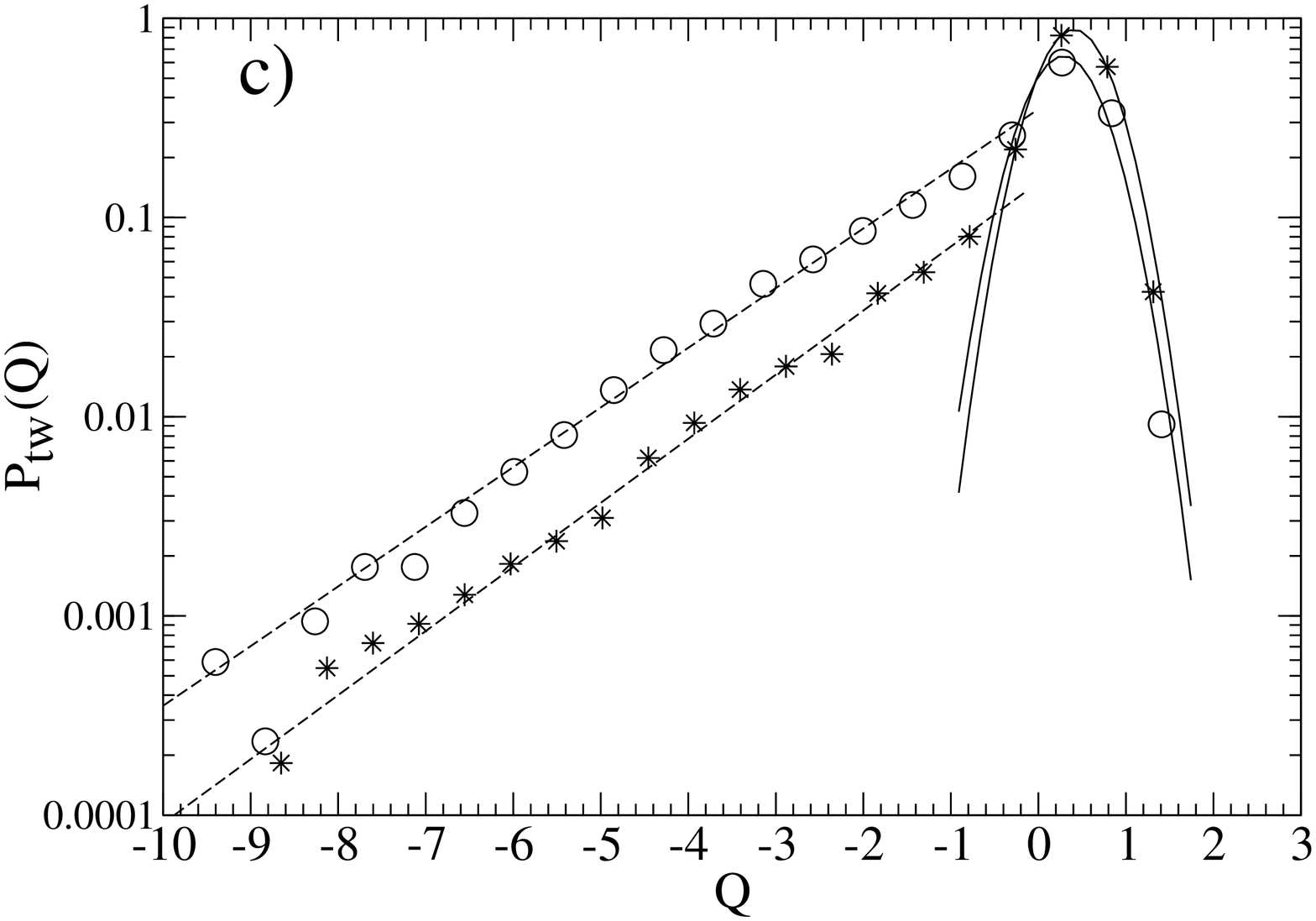,angle=-90, width=9cm}
\end{center}
\caption{Heat exchange PDFs for $T=0.3$ (panel a),
$T=0.2$ (panel b), $T=0.1$ (panel c). Circles are for $t_w=2^{10}$ and
asterisks for $t_w=2^{15}$. The continuous lines are Gaussian fits to the stimulated sector,the
dashed lines are the exponential fits to the spontaneous sector. The parameters of the fits are given in Table~\protect\ref{table}.}
\label{fig2}
\end{figure}

Figure~\ref{fig2} shows the $P_{t_w}(Q)$ for different values of $t_w$
and $T$. It clearly shows the existence of two sectors, a Gaussian
sector for small heat values of $Q$ and an exponential tail extending
down to negative $Q$ values. A salient feature of this figure is the
clear cut distinction existing between the two sectors, the total
amount of heat released by the spontaneous mechanism is generally
quite small (compared to the overall absolute value of the heat
exchanged through the stimulated process) yet the spontaneous decay is
the leading mechanism by which heat is released to the bath. The
stimulated and the spontaneous sectors can be very well fitted to a
Gaussian $P_{t_w}^{\rm st}(Q)\sim \exp(-(Q-a)^2/(2b^2))$ and an
exponential function $P_{t_w}^{\rm sp}(Q)\sim \exp(Q/\lambda)$
respectively, the normalization constant of these distributions being
unimportant. In general there is no reason a priori for the
spontaneous component not to display a Gaussian correction,
$P_{t_w}^{\rm sp}(Q)\sim \exp(Q/\lambda+{\cal O}(Q^2))$, yet in the
present case this correction appears negligible.  In Table I we show
the results obtained for the fitting parameters.  The numbers there
reported confirm the scenario previously described.  As the
temperature of the bath decreases the width of the Gaussian $b^2$
decreases. On the other hand, the width of the Gaussian is nearly
$t_w$ independent and the width of the exponential tail $\lambda$
decreases with $t_w$.  Note that the average heat
exchanged through the stimulated process (the parameter $a$) is
different form zero, the reason being our protocol where values of $Q$
are measured along non-regularly spaced time intervals.
\begin{table}[b]
 \begin{center}
\begin{tabular}{|c|c|c|c|c|}\hline
$T$ & $t_w$ & $a$ & $b^2$ & $\lambda$  \\\hline
 0.3 & $2^{10}$ &0.98 & 1.03 & 0.77\\\hline
0.3 & $2^{15}$  & 1.34 & 1.04 & 0.62 \\\hline
0.2 & $2^{10}$ & 0.75 & 0.50 & 1.00 \\\hline
0.2 & $2^{15}$  & 0.88 & 0.49 & 0.90\\\hline
0.1 & $2^{10}$ & 0.29 & 0.17 & 1.45\\\hline
0.1 & $2^{15}$ & 0.40 & 0.16 & 1.35\\\hline
 \end{tabular}
 \caption{Fit parameters for the Gaussian and exponential fits of Fig.~\protect\ref{fig2}}
\label{table}
 \end{center}
\end{table}

\begin{figure}
\begin{center}
\epsfig{file=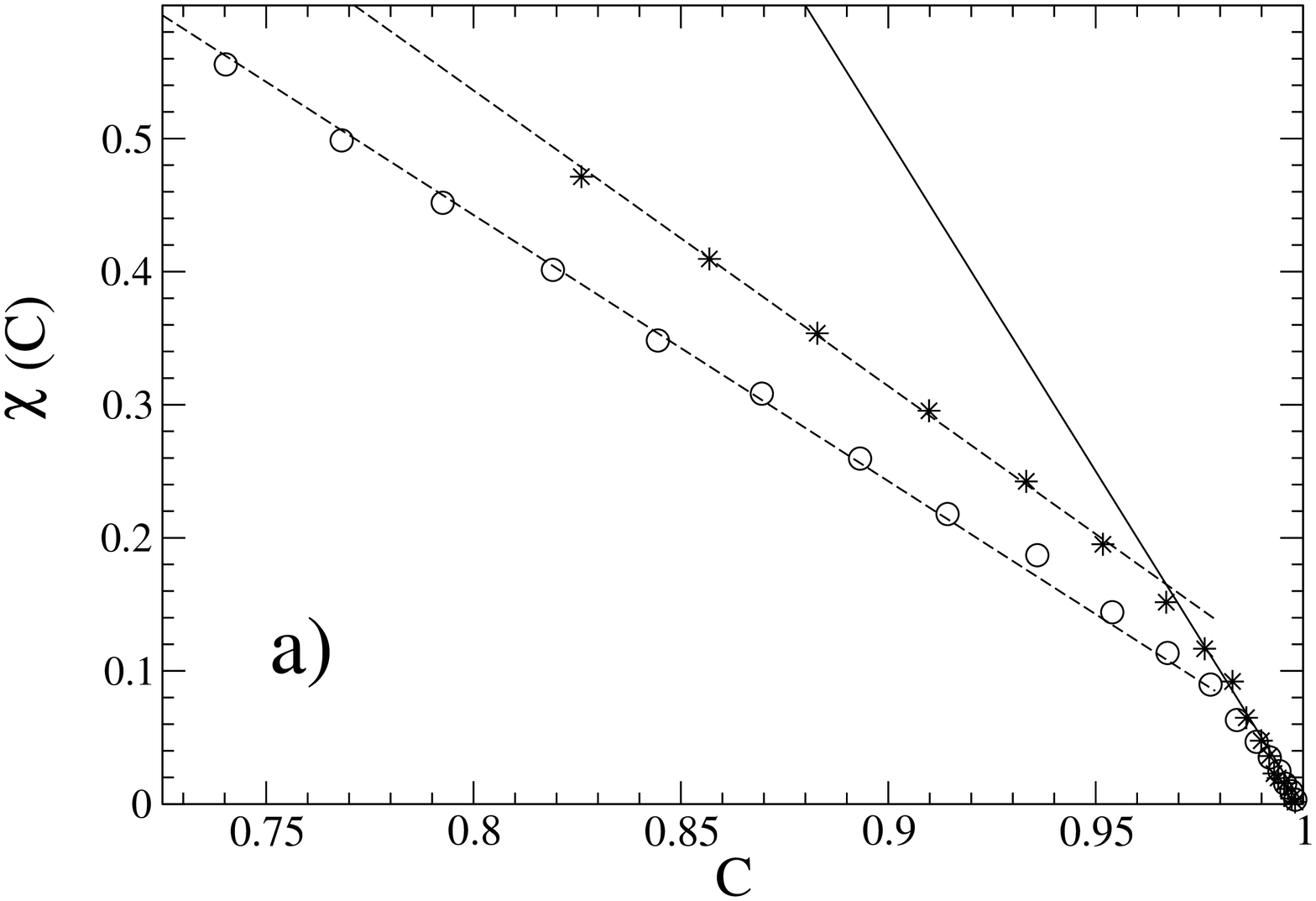,angle=-90,width=9cm}\epsfig{file=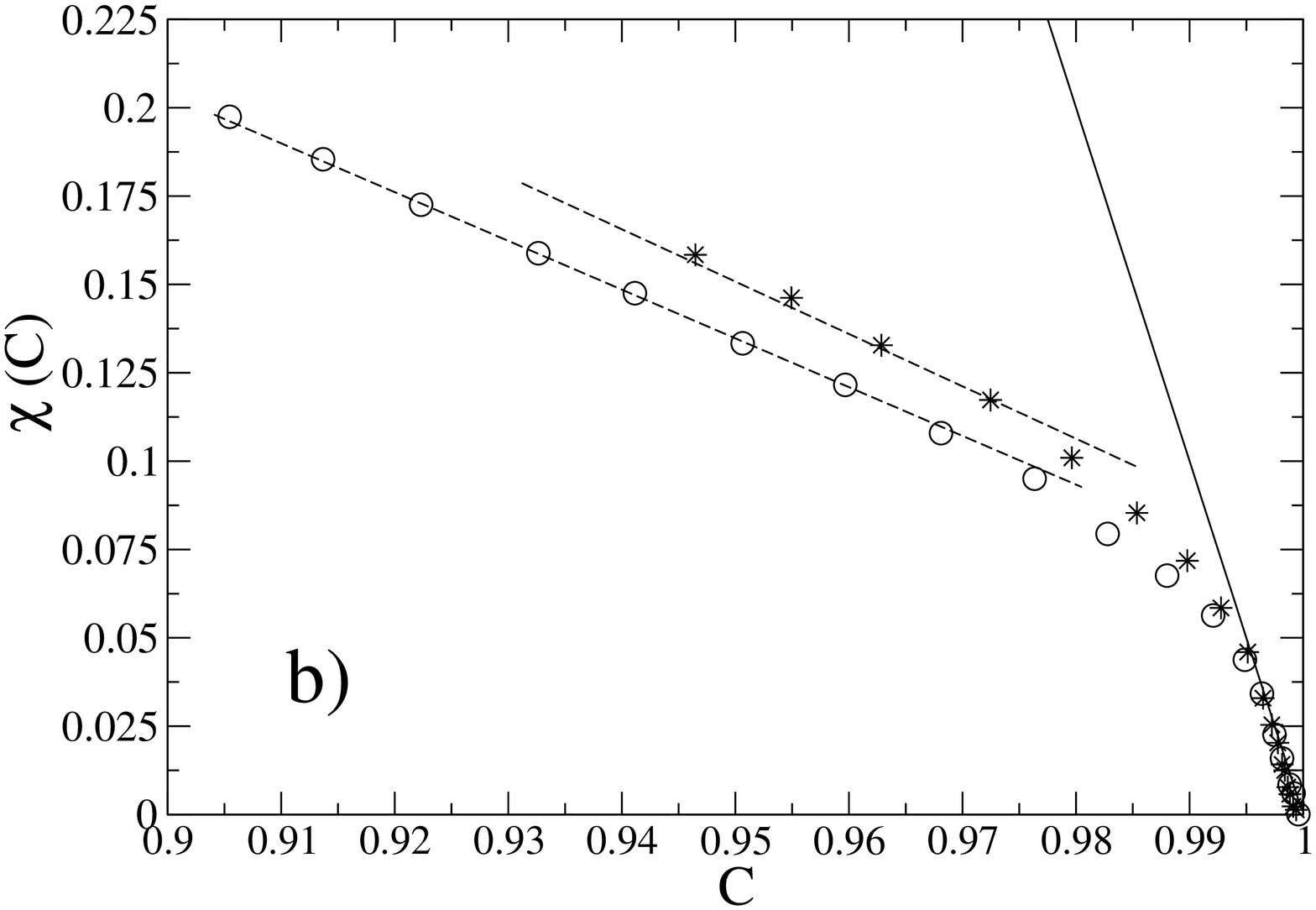,angle=-90, width=9cm}
\end{center}
\caption{Fluctuation-dissipation plots for $T=0.2$ (panel a), $T=0.1$
(panel b) and $t_w=2^{10}$ (circles) and asterisks for $t_w=2^{15}$
(asterisks). The intensity of the field is $h_0=0.1$. The continuous
line is the equilibrium result $\chi(C)=(1-C)/T$, the dashed lines are
single parameter fits to the linear relation $\chi(C)=a+C/T_{\rm eff}$
where $a$ is the fit parameter and $T_{\rm eff}=\lambda/2$ is obtained
from the values reported in Table~\protect\ref{table}.}
\label{fig4}
\end{figure}

The result for the spontaneous component $P^{\rm sp}_{t_w}(Q)\propto
\exp(Q/\lambda+{\cal O}(Q^2))$ can be recast in the form of a
FT~\cite{EvaSea02},
\be
\frac{P^{\rm sp}_{t_w}(Q)}{P^{\rm sp}_{t_w}(-Q)}=\exp\Bigl
(\frac{2Q}{\lambda }\Bigr)~~~~.
\label{a4}
\ee
An explicit numerical check of this identity requires to identify the
spontaneous component $P^{\rm sp}_{t_w}(Q)$ out of the global
distribution $P_{t_w}(Q)$. This is a difficult task, as heat
fluctuations with $Q>0$ are masked by the stimulated
component. Actually spontaneous events with $Q>0$ are never observed,
however they enter into the formulation of the FT. We plan to
substantiate the fact that the easiest way to probe spontaneous
transitions with $Q>0$ is by applying an external perturbation that
lifts the energy levels of the system making these transitions
accessible. This is accomplished by the evaluation of
fluctuation-dissipation relations as they specifically contain these
transitions.  A general feature of glassy systems is the existence of
violations of the fluctuation-dissipation theorem
(FDT)~\cite{CriRit03} that lead to a modified version of the theorem
and have been interpreted in terms of an effective macroscopic
temperature $T_{\rm eff}$~\cite{CugKurPel98}.  An important aspect of
effective temperatures is that they are generally uncoupled to the
temperature of the bath, their emergence thought to be related to the
presence of an exponential density of states.  Previous considerations
and the validity of \eq{a4} call for a connection between the value of
$\lambda$ and the effective temperature $T_{\rm eff}$ as derived from
the modified FDT. The outcome of all these considerations is that
$\lambda\simeq 2T_{\rm eff}$, the factor $2$ being consequence of the
validity of the FT \eq{a4}. The proof of this relation is based on the
presence of an exponential tail of the heat released from system to
bath and a microcanonical entropic argument {\it a la Edwards} used in
granular media \cite{Edwards90} that counts the number of valleys with
free energy $F$ available to the system (for a discussion of how
partitioning of the phase space in valleys can be accomplished see
\cite{CriRit03}). Let $\Omega(F)$ stand for the number of valleys of
free-energy $F$ and let $Q$ be the heat released to the system when
jumping from a valley of free energy $F$ to a valley of free energy
$F'$. Because spontaneous transitions are entropically driven the
distribution $P^{\rm sp}_{t_w}(Q)$ is proportional to the number
$\Omega(F')$. For the ratio between the forward and reverse
transitions we can write,
\be
\frac{P^{\rm sp}_{t_w}(Q)}{P^{\rm
sp}_{t_w}(-Q)}=\frac{\Omega(F')}{\Omega(F)}~~~~.
\label{a1}
\ee
The number of valleys with a given free energy defines a configurational
entropy of valleys $S_c(F)=\log(\Omega(F))$. Inserting this dependence
and identifying \eq{a1} with \eq{a4} we obtain,
\be
\exp(\frac{2Q}{\lambda})=\exp\Bigl(\frac{\partial S_c(F)}{\partial F}\Delta
F\Bigr)
\label{a2}
\ee
where $\Delta F=F'-F$ and where we have kept  the linear term in $\Delta F$ in
the r.h.s. Using the relation $\Delta F=\Delta E-T\Delta S$
and the identity $Q=\Delta E$ as well as the definition for
the effective temperature $1/T_{\rm eff}=\frac{\partial S_c(F)}{\partial
F}$ we obtain the result,
\be
\frac{1}{\lambda}=\frac{1}{2T_{\rm eff}}\bigl( 1-T\frac{\partial S(E)}{\partial E} \bigr) 
\label{a3}
\ee
where $S(E)$ denotes the dependence of the entropy of individual
valleys on their average energy $E$. A good approximation amounts to
retain valleys as having very similar properties, in that case $S(E)$
is only a function of the temperature of the bath and $\lambda=2T_{\rm
eff}$. In other cases, however, $S(E)$ is simply very small and so is
the term $\frac{\partial S(E)}{\partial E}$. This is the case of the
ROM where, among other features, the entropy of valleys has been shown
to be quite small~\cite{Coluzzi03} so the identity $\lambda=2T_{\rm
eff}$ is a very good approximation. Let us mention that both $\lambda$
and $T_{\rm eff}$ in \eq{a3} are time dependent, an indirect
manifestation of the existence of activated processes.  The identity
\eq{a3} is challenged in Figure~\ref{fig4} were we show the
corresponding fluctuation-dissipation plots. These are constructed by
evaluating the zero-field cooled susceptibility $\chi(t,t_w)$ at time
$t$ after applying a small field $h_0$ at $t_w$, the corresponding
correlation $C(t,t_w)$, and plotting one in terms of the
other~\cite{CriRit03}.  The identity $\lambda=2T_{\rm eff}$ is well
verified in all cases.

To better justify the FT \eq{a4} and the identity
\eq{a3} we have considered a family of exactly solvable models where
these results can be explicitly checked. The family of models is defined
by an ensemble of $N$ one-dimensional oscillators described by the
variables $\lbrace x_i; 1\le i\le N\rbrace$ where $-\infty<x_i\le
\infty$. Oscillators are subject to a potential
$V(x)=\frac{k}{2p}x^{2p}$ where $k$ is a stiffness constant and $p$ is
an integer $p\ge 1$. The total energy of the ensemble is given by
$E=\sum_{i=1}^NV(x_i)$.  The model being non-interacting has no phase
transition at finite temperature and the partition function is given by
${\cal Z}={\cal Z}_1^N$ with ${\cal
Z}_1=\int_{-\infty}^{\infty}dx\exp(-\beta V(x))$.  For this family of
models to show glassy behavior at low temperatures we consider a Monte
Carlo dynamics where oscillator positions are updated in parallel
$x_i\to x_i+\frac{r_i}{\sqrt{N}}$, and the $r_i$ are random distributed
variables of zero mean and variance $\Delta^2$. The case $p=1$
corresponds to the harmonic oscillator model introduced in
\cite{BonPadRit98}. The Monte Carlo dynamics generates glassy behavior
at $T=0$ as the ground state configuration $x_i=0$ has zero measure,
thereby remaining unaccessible due to the finite value of the step
distance $\Delta$. At $T=0$ relaxation becomes logarithmic and the
system partially equilibrates over the surface of constant energy with
spontaneous energy decays whenever a configuration with lower energy is
found.  In such adiabatic regime each configuration corresponds to a
valley so the identity $\lambda=2T_{\rm eff}$ is expected to be
exact. Moreover because at $T=0$ there is no stimulated component, all
heat exchanges are spontaneous and $P_{t_w}(Q)=P_{t_w}^{\rm
sp}(Q)$. Rather than addressing the full dynamical solution
\cite{Ritort03} we content ourselves to present the main steps of the
calculation.  To compute the $P_{t_w}(Q)$ is enough to determine the PDF
of energy changes given by $\Delta E=Q
=\frac{k}{2p}\sum_i\sum_{l=1}^{2p}{2p\choose l}
x_i^{2p-l}r_i^lN^{-l/2}$.  It is easy to verify that such a distribution
is a Gaussian of mean $\overline{Q}=\frac{k(p-1)}{2}\Delta^2 h_p$ and
variance $\overline{Q^2}-\overline{Q}^2=k^2\Delta^2 h_{2p}$ with
$h_p=\langle x^{2(p-1)}\rangle$, the average taken over dynamical
histories. Inserting the expression for the Gaussian in the FT \eq{a4}
gives the exact result $\lambda=\frac{2k h_{2p}}{(2p-1)h_p}$. For
relaxation is adiabatic, the moments $h_p$, albeit time dependent
functions through the energy $E$, are related each other by the
corresponding equilibrium relations. A straightforward calculation gives
$h_{2p}/h_p=(2p-1)(2p)E/Nk$. Inserting this expression into the previous
one we get $\lambda=4pE/N$, and using the relation
$S_c(E)=\frac{N}{2p}\log(E)$ we obtain $\lambda=2/S_c'(E)=2T_{\rm
eff}$. As we said, the total entropy coincides with the configurational
entropy as configurations correspond to valleys.

Albeit in a different context, a FT strikingly similar to \eq{a4} has
been recently obtained~\cite{ZonCoh03} for the case of a Brownian
particle in a steady state when dragged by a moving harmonic potential
and subjected to a viscous drag force. The result \eq{a4} underlies
the connection between intermittency and entropy production in the
aging regime beyond the case of non-equilibrium systems in their
steady states~\cite{Gal00,Sel98}.  This raises the intriguing
possibility that heat exchange FTs are widespread in condensed
matter physics in many non-equilibrium situations.

The experimental determination of the two types of heat emission here
described could be addressed in Nyquist noise measurements of glass
formers where voltage noise fluctuations induce heat dissipation in
the resonant cavity. Indeed the heat dissipated by a resonant cavity
of impedance $Z(\omega)$ should be proportional to
$V^2(\omega)/Z(\omega)$ and intermittent bursts in the voltage signal
would correspond to spontaneously dissipated heat events. Actually,
recent experiments~\cite{Ciliberto03} have obtained PDFs of the
voltage signal $V$ whose profile is strikingly similar to the results
shown in Figure~\ref{fig2}. Moreover, these profiles show the presence
of exponential tails with a slowly decaying width (therefore, in
agreement with our results) that we interpret in terms of a
time-dependent effective temperature. Conversion of the voltage into
dissipated power will result in a signal whose PDF still shows
intermittency. The width of the exponential tail will satisfy
$\lambda(t_w)=c T_{\rm eff}(t_w)$, the structural constant $c$
depending on the cavity as well as on the type of coupling between the
system and the cavity. Similarly, the non-Gaussian tails observed by
time resolved correlation analysis of intensity signals in colloidal
glasses \cite{BisTraRomCip03} could be interpreted in terms of
exponential tails describing spontaneous events where heat is released
to the bath due to the existence of CRRs.  In general, intermittency
measurements offer a potentially interesting vein where new
quantitative results on the non-equilibrium dynamics of slow systems
can be inferred.  Finally, let us mention that the existence of a FT
that governs heat exchanges in the non-equilibrium aging state can be
further investigated in numerical and theoretical investigations of
other models for the glass transition. Preliminary
investigations in Lennard-Jones binary mixtures confirm all the
results we have presented for the ROM.  A detailed exposition of these
results will be presented in the future. These ideas could be also
extended to granular systems where now relaxed free volume (instead of
released heat) would describe the relaxation~\cite{BruRei03}. The
results here purported suggest the existence of an interesting link
between general theoretical aspects of the non-equilibrium dynamics of
glass formers in their aging regime and intermittent noise signals
experimentally measurable by different methods. Further work in
experiments, theory and simulations will clarify the implications and
potentialities of the present approach.

{\bf Acknowledgments.} FR is supported by the Spanish Ministerio de
Ciencia y Tecnolog\'{\i}a Grant BFM2001-3525 and Generalitat de
Catalunya.

\end{document}